\documentclass[twocolumn,preprintnumbers,amsmath,amssymb]{revtex4}

\usepackage{graphicx}
\usepackage{dcolumn}
\usepackage{bm}
\newcommand {\bsub} {\begin{subequations}}
\newcommand {\esub} {\end{subequations}}
\newcommand {\bea} {\begin{eqnarray}}
\newcommand {\eea} {\end{eqnarray}}

\begin{document}

\preprint{OCHA-PP-231}

\title{
Search for a light extra gauge boson in Littlest Higgs 
model at a linear collider
}

\author{Gi-Chol Cho}
\affiliation{%
Department of Physics, Ochanomizu University,\\
Tokyo, 112-8610, Japan
}%
\author{Aya Omote}
\affiliation{%
Graduate School of Humanities and Sciences, \\
Ochanomizu University, Tokyo, 112-8610, Japan
}%

\begin{abstract}
Littlest Higgs model predicts some extra particles beyond the 
Standard Model. 
Among them, an extra neutral gauge boson $A_H$ is lightest and 
its mass could be a few hundred GeV. 
We study production and decay of $A_H$ at future $e^+ e^-$ linear 
collider and compare them with those of $Z'$ bosons in supersymmetric 
(SUSY) $E_6$ models. 
We find that, if the extra gauge boson mass is smaller than $\sqrt{s}$ 
of the linear collider, the forward-backward asymmetries of $b$- and 
$c$-quarks at the $A_H$ pole differ significantly from those given by 
the $Z'$ bosons, and are useful to test the littlest Higgs model and SUSY 
$E_6$ models. 
\end{abstract}

\maketitle
Little Higgs~\cite{Arkani-Hamed:2001nc, Arkani-Hamed:2002pa,
Arkani-Hamed:2002qx} 
is an attractive idea to solve the gauge hierarchy problem. 
In this class of models, the electroweak Higgs boson appears as 
a pseudo-goldstone boson of a certain global symmetry breaking 
at a scale $\Lambda \sim 10~{\rm TeV}$ 
so that the Higgs boson mass can be as light as $O(100~{\rm GeV})$. 
The light Higgs boson mass is protected from the 1-loop quadratic
divergence by gauging a part of global symmetry, and introducing 
a few extra heavy particles whose typical mass scale is of order 
$f \equiv \Lambda/4\pi$, where $f$ is a decay constant of the 
pseudo-goldstone boson. 
%

%
One of the simplest models based on this idea is known as Littlest Higgs 
Model~\cite{Arkani-Hamed:2002qy}, which has a global SU(5) and 
gauged $[{\rm SU(2)}\times {\rm U(1)}]^2$ symmetries at ultra-violet regime. 
At the scale $\Lambda$, the global symmetry is broken to SO(5) while 
the gauge symmetry is broken to that of the Standard Model (SM),  
SU(2)$_L\times$U(1)$_Y$. 
Such symmetry breaking allows that the littlest Higgs model has four
massive gauge bosons 
below $\Lambda$, and they are mixed with the SM gauge bosons after 
the electroweak symmetry breaking. 
As a result, the set of extra gauge bosons at the weak scale consists of
electrically neutral states ($A_H, Z_H$) and charged states 
($W_H^\pm$). 
The mass scale of the extra gauge bosons are given by 
$m_{A_H} \sim g_Yf$ and $m_{Z_H} \approx m_{W_H} \sim g f$, where 
$g_Y$ and $g$ are U(1)$_Y$ and SU(2)$_L$ gauge couplings, respectively. 
In addition to the extra gauge bosons, the littlest Higgs model predicts 
a triplet scalar $\Phi$ and a vector-like quark $T$ in order to 
stabilize the Higgs boson mass against the radiative corrections. 
The scalar $\Phi$ cancels the 1-loop quadratic divergence by the Higgs 
self-interaction, while the extra quark $T$ cancels one by the top-quark
Yukawa interaction. 
Up to the order one coefficients, the masses of $\Phi$ and $T$ are
roughly given by the decay constant $f$~\cite{Han:2003wu}.  
Among the extra particles in the littlest Higgs model, therefore, 
$A_H$ is lightest one, so that it is expected to be discovered at future 
collider experiments rather early. 
In hadron collider experiments, $A_H$ is produced in the Drell-Yan 
process, $p\bar{p}~({\rm or~}pp) \to A_H \to \mu^+ \mu^- X$, and shows 
a peak of the invariant mass distribution of muon pair in the final 
state~\cite{Han:2003wu, Hewett:2002px}. 
%

%
Since such an experimental signature of the $A_H$ boson is quite similar 
to a $Z'$ boson in models which have an extra U(1) gauge symmetry, 
it is very important to identify the models if an extra neutral gauge 
boson is discovered in hadron collider experiments such as Tevatron
Run-II or LHC. 
The discovery limit of the $Z'$ boson at the LHC experiment is expected
to be roughly 3 TeV~\cite{Golutvin:2003gc}. 
However, it is hard to test the $Z'$ models at the hadron colliders. 
On the other hand, an $e^+ e^-$ linear collider can play a 
complementary role for such purpose~\cite{Aguilar-Saavedra:2001rg}. 
In this paper, we would like to study production and decay of 
$A_H$ in the littlest Higgs model at a future $e^+ e^-$ linear collider.  
In particular, we compare the experimental signatures of $A_H$ and  
those of $Z'$ boson in supersymmetric $E_6$ models~\cite{Hewett:1988xc}, 
and examine a possibility to distinguish these models in the linear 
collider experiments. 
In our study, we assume that Tevatron or LHC discovers a certain $Z'$ 
boson whose mass is smaller than $\sqrt{s}$ of the linear collider. 
Then we can study the $Z'$ boson at the linear collider by tuning 
the $e^+ e^-$ beam energy at the peak of $Z'$ resonance. 
We will show that the forward-backward asymmetry of $c$-quark 
is useful to test if the $Z'$ boson is $A_H$ in the littlest Higgs model 
or one of the SUSY $E_6$ models. 
%

%
Since the littlest Higgs model has the gauge symmetry 
$\left[{\rm SU(2)} \times {\rm U(1)}\right]^2$ at high energy scale, 
there are two SU(2) gauge bosons $W_1, W_2$ and 
U(1) gauge bosons $B_1, B_2$. 
The global SU(5) symmetry breaking at the scale $\Lambda$ induces the
following mixing among them: 
\begin{eqnarray}
\left(
\begin{array}{c}
W^a \\ W'^a 
\end{array}
\right) &=& \left(
\begin{array}{rr}
s_\theta & c_\theta \\
-c_\theta & s_\theta
\end{array}
\right)
\left(
\begin{array}{c}
W_1^a \\ W_2^a 
\end{array}
\right), 
\\
\left(
\begin{array}{c}
B \\ B' 
\end{array}
\right) &=& \left(
\begin{array}{rr}
s_{\theta'} & c_{\theta'} \\
-c_{\theta'} & s_{\theta'}
\end{array}
\right)
\left(
\begin{array}{c}
B_1 \\ B_2
\end{array}
\right), 
\label{thetaprime}
\end{eqnarray}
where $W^a$ and $B$ are SU(2)$_L$ and U(1)$_Y$ gauge bosons in the SM, 
respectively. They are massless at this stage while $W'^a$ and $B'$ are 
massive. 
Note that we use the shorthand notation 
$s_\theta \equiv \sin\theta$, 
$c_\theta\equiv \cos\theta$,
$s_{\theta'} \equiv \sin\theta'$, 
$c_{\theta'}\equiv \cos\theta'$. 
After the electroweak symmetry is broken, the massless gauge bosons 
will acquire masses and be mixed with $W'^a$ and $B'$. 
The mass eigenstates for charged and neutral gauge bosons are obtained 
by introducing unitary matrices $U_W$ and $U_N$ as: 
\begin{eqnarray}
\left(
\begin{array}{c}
W_L \\ W_H 
\end{array}
\right) &=& U_W 
\left(
\begin{array}{c}
W \\ W' 
\end{array}
\right), 
\\
\left(
\begin{array}{cccc}
A_L & Z_L & A_H & Z_H
\end{array}
\right)^{\rm T} &=& U_N 
\left(
\begin{array}{cccc}
B & W^3 & B' & W^{'3}
\end{array}
\right)^{\rm T}. 
\end{eqnarray}
The explicit expressions of the unitary matrices $U_W, U_N$ and the mass 
eigenvalues of gauge bosons can be found in ref.~\cite{Han:2003wu}. 
%

%
The interaction of $A_H$ to a fermion $f$ in the SM is described by the 
following Lagrangian~\cite{Han:2003wu}: 
\begin{eqnarray}
{\cal L} &=& -\frac{g_Y}{2 s_{\theta'}c_{\theta'}}
Q_{f_\alpha}^{A_H} \overline{f_\alpha} \gamma^\mu f_\alpha 
A_{H\mu}, 
\label{interaction}
\end{eqnarray}
where $\alpha(=L,R)$ denotes the chirality of fermion $f$. 
We summarize the charge $Q^{A_H}_{f_\alpha}$ for the fermion $f_\alpha$ 
in Table~\ref{lhm_charge}, which is obtained by taking account of the 
$\left[{\rm SU}(2)\times {\rm U(1)}\right]^2$ anomaly cancellation
without introducing any chiral fermions beyond the SM~\cite{Han:2003wu}. 
\begin{table}[h]
\begin{center}
\begin{tabular}{c|ccccc} \hline 
field & 
$L$ & $e_R$ & $Q$ & $u_R$ & $d_R$ \\ \hline
charge &  $-\frac{2}{5} + c_{\theta'}^2$
& $-\frac{4}{5} + 2c_{\theta'}^2$ 
& $\frac{2}{15} - \frac{1}{3}c_{\theta'}^2$ 
& $\frac{8}{15} - \frac{4}{3}c_{\theta'}^2$ 
& $-\frac{4}{15} + \frac{2}{3}c_{\theta'}^2$  \\ \hline
\end{tabular}
\end{center}
\caption{
Charge $Q_{f_\alpha}^{A_H}$ for quarks and leptons. 
$L$ and $Q$ denote the SU(2)$_L$ doublet lepton and quark, 
respectively. 
The charges are determined to satisfy the 
$\left[{\rm SU}(2)\times {\rm U(1)}\right]^2$ anomaly free condition 
without introducing any chiral fermions beyond the SM. 
}
\label{lhm_charge}
\end{table}
%

%
Next let us briefly review the supersymmetric $E_6$ models to fix our 
notation~\cite{Hewett:1988xc}. 
Since the rank of $E_6$ is six, it has two U(1) factors besides the SM
gauge group which arise from the following decompositions: 
\begin{eqnarray}
        \begin{array}{rl}
        E_6 &\supset {\rm SO(10)} \times {\rm U(1)}_\psi
\\
        &\supset {\rm SU(5)} \times {\rm U(1)}_\chi \times {\rm U(1)}_\psi.
        \end{array}
\end{eqnarray}
An additional $Z'$ boson in the electroweak scale can be parametrized as
a linear combination of the U(1)$_\psi$ gauge boson $Z_\psi$ and 
the U(1)$_\chi$ gauge boson $Z_\chi$ as 
\begin{eqnarray}
Z' &=& Z_\psi \cos\beta_E + Z_\chi \sin\beta_E. 
\end{eqnarray}
There are four $Z'$ models, which are called $\chi,\psi,\eta$ and $\nu$
models, corresponding to the different value of the mixing angle 
$\beta_E$. 
The interaction of $Z'$ boson with the fermion $f$ is described as: 
\begin{eqnarray}
{\cal L} &=& - g_E Q_E^{f_\alpha}  \overline{f_\alpha} \gamma^\mu f_\alpha 
Z_\mu'
\end{eqnarray}
where $g_E$ denotes the extra U(1) gauge coupling constant. 
The extra U(1) charge $Q_E^{f_\alpha}$ for the SM quarks and leptons 
in four $Z'$ models are summarized in Table~\ref{charge_e6}. 
\begin{table}[h]
\begin{center}
\begin{tabular}{|c||c|c|c|c|c|} \hline
field & $Y$ & $2\sqrt{6} Q_\chi$ & $\sqrt{72/5}Q_\psi$ & $6 Q_\eta$ 
& $\sqrt{6}Q_\nu$ \\ \hline 
$Q$ & $1/6$ & $-1$ & $1$ & $-2$ & $1/2$ \\
$u_R$ & $2/3$ & $1$ & $-1$ & $2$ & $-1/2$ \\
$e_R$ & $-1$ & $1$ & $-1$ & $2$ & $-1/2$ \\
$L$ & $-1/2$ & $3$ & $1$ & $1$ & $1$ \\
$d_R$ & $-1/3$ & $-3$ & $-1$ & $-1$ & $-1$ \\ \hline
\end{tabular}
\end{center}
\caption{
The hypercharge $Y$ and the extra U(1) charge $Q_E$ of SM quarks and leptons 
in SUSY $E_6$ models. The value of extra U(1) charge follows the 
hypercharge normalization. 
}
\label{charge_e6}
\end{table}
%

%
It should be noted that not only in the littlest Higgs model but also 
in the SUSY $E_6$ models, 
the extra neutral gauge boson can mix with the SM $Z$ boson after 
the electroweak symmetry breaking. 
Such mixing is, however, severely constrained from the experimental data 
of the electroweak precision measurements at the $Z$-pole~\cite{Cho:1998nr}, 
and we assume that the mixing is small enough to neglect in our study. 
%

%
\begin{figure}[ht]
\includegraphics[width=8cm,clip]{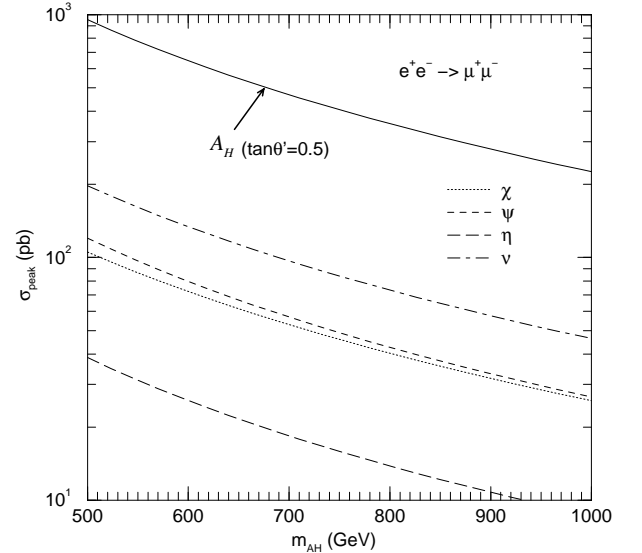}
\caption{The peak cross section of $e^+ e^- \to \mu^+\mu^-$ at the $A_H$ 
or $Z'$ pole in the littlest Higgs model (solid line) and SUSY $E_6$ 
models as functions of $A_H$ (or $Z'$) mass. 
The mixing angle $\theta'$ in the littlest Higgs model is fixed by 
$\tan\theta'=0.5$. 
}
\label{fig:peak}
\end{figure}
Now let us examine theoretical predictions of extra gauge boson in each model 
at the linear collider. 
As is already mentioned, we assume that the mass of $A_H (Z')$ is 
smaller than $\sqrt{s}$ of the linear collider, i.e., 
we expect that one can measure observables in $e^+e^-\to f\bar{f}$ 
processes at the peak of $A_H(Z')$ resonance. 
The peak cross section of $e^+ e^- \to f \bar{f}$ is given by 
(note $V=A_H$ or $Z'$)
\begin{eqnarray}
\sigma_{\rm peak}^f &=& \frac{12\pi}{m_V^2}
\frac{\Gamma_e \Gamma_f}{\Gamma_V^2}, 
\label{peak_cross_section}
\\
\Gamma_f &=& \frac{m_V}{24\pi}
\left\{\left(|g_L^f|^2+|g_R^f|^2\right)
       \left(1-\frac{m_f^2}{m_V^2}\right)
\right.
\nonumber \\
&&\left.
+ 6\frac{m_f^2}{m_V^2} {\rm Re}(g_L^f g_R^{f*})
\right\}
\label{width_a}
\end{eqnarray}
where $\Gamma_f$ is the partial decay width of $V\to f \bar{f}$ 
and $\Gamma_V$ is the total decay width of $V$. 
The coupling $g_\alpha^f (\alpha=L,R)$ in (\ref{width_a}) follows the 
normalization
\begin{eqnarray}
{\cal L} &=& -g_\alpha^f \overline{f_\alpha}\gamma^\mu f_\alpha V_\mu. 
\label{normalization}
\end{eqnarray}
We note that $\sigma^f_{\rm peak}$ does not depend on the gauge 
coupling ($g_Y$ in the littlest Higgs model, $g_E$ in SUSY $E_6$ models)
because it is canceled in (\ref{peak_cross_section}). 
In Fig.~\ref{fig:peak}, we show the peak cross section of 
$e^+e^- \to \mu^+\mu^-$ in the littlest Higgs model and SUSY $E_6$ models 
as a function of extra gauge boson mass $m_{A_H}(m_{Z'})$. 
The prediction of the littlest Higgs model is shown for $\tan\theta'=0.5$ 
as an example. 
The peak cross section in the littlest Higgs model is roughly a few
hundred pb, which is a few times larger than those in SUSY $E_6$ models, 
so the cross section measurement seems to be useful to test the models 
at the linear collider with $100 {\rm fb}^{-1}$ integrated luminosity.  
%

%
\begin{figure}[ht]
\includegraphics[width=8cm,clip]{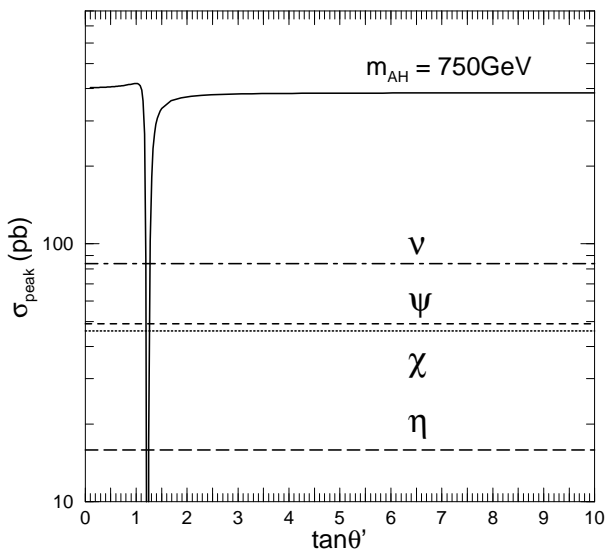}
\caption{
The peak cross section of $e^+e^- \to \mu^+\mu^-$ in the littlest 
Higgs model (solid line) and SUSY $E_6$ models as functions of 
the mixing angle $\theta'$. The mass of $A_H$ (or $Z'$) is fixed at 
750 GeV. 
}
\label{fig:angle}
\end{figure}
Next we show the peak cross section of $e^+e^- \to \mu^+\mu^-$ in 
the littlest Higgs model as a function of the mixing angle $\theta'$ 
in Fig.~\ref{fig:angle}.  
The mass of $A_H$ is fixed at 750 GeV as an example. 
For comparison, we depict the predictions of SUSY $E_6$ models in the 
same figure. 
We can see in the figure that the peak cross section in the littlest 
Higgs model rapidly decreases around $\tan\theta' = 1.2$. 
This is because that the left- and right-handed electron couplings to 
$A_H$ are given by (see Table~\ref{lhm_charge})
\begin{eqnarray}
Q^{A_H}_{e_L} = \frac{1}{2} Q^{A_H}_{e_R} 
= \frac{1}{2 s_{\theta'} c_{\theta'}} \left(-\frac{2}{5}+c_{\theta'}^2
\right), 
\label{mixing} 
\end{eqnarray}
and they diminish for $c_{\theta'}^2 \sim 2/5$, which corresponds 
to $\tan\theta'\sim 1.2$. 
Therefore, even if the result of peak cross section measurement is
consistent with one of SUSY $E_6$ models, there is still a possibility 
of the littlest Higgs model, and we should find another observable to
test the models. 
%

\begin{figure}[ht]
\includegraphics[width=8cm,clip]{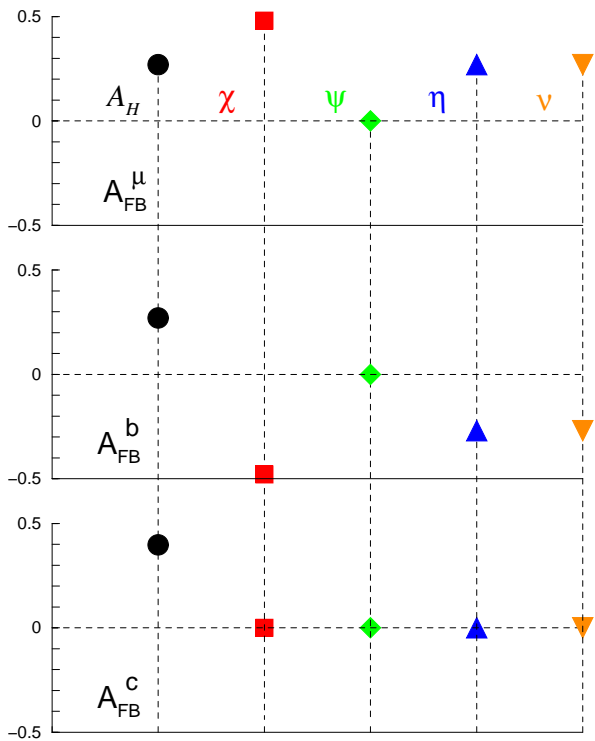}
\caption{
The forward-backward asymmetry of muon (top), $b$-quark (middle) and 
$c$-quark (bottom) at the $Z'$ pole in the littlest Higgs model and 
SUSY $E_6$ models. 
The predictions of littlest Higgs model is shown by solid circles, while 
those of SUSY $E_6$ models are given by squares ($\chi$-model), 
diamonds ($\psi$-model), upward triangles ($\eta$-model) and 
downward triangles ($\nu$-model). 
}
\label{fig:asymmetry}
\end{figure}

%
The forward-backward (FB) asymmetry of the $e^+ e^- \to f \bar{f}$ 
process does not have the $\theta'$ dependence. 
Using the couplings defined in (\ref{normalization}), 
the FB asymmetry at the pole of $A_H(Z')$ can be expressed as 
\begin{eqnarray}
A_{\rm FB}^f &=& \frac{3}{4} A^e A^f, \\
A^f &=& \frac{(g_L^f)^2 - (g_R^f)^2}{(g_L^f)^2 + (g_R^f)^2}. 
\label{eq_asym}
\end{eqnarray}
When we write the right-handed coupling $g_R^f$ as 
\begin{eqnarray}
g_R^f \equiv r_f g_L^f, 
\end{eqnarray}
the asymmetry parameter $A^f$ is expressed as follows 
\begin{eqnarray}
A^f = \frac{1-r_f^2}{1+r_f^2}. 
\end{eqnarray}
In the littlest Higgs model, the coupling $g^f_\alpha$ is replaced by 
$g_Y Q^{A_H}_{f_\alpha}/(2s_{\theta'}c_{\theta'})$ (see 
eq.(\ref{interaction})), so the parameter $r_f$ for $f=e,u,d$ 
is independent of $\theta'$; 
\begin{eqnarray}
(r_e,r_u,r_d)=(2,4,-2). 
\end{eqnarray}
The FB asymmetry, therefore, is a good observable to compare the 
littlest Higgs model and the SUSY $E_6$ models. 
We show the FB asymmetry for the muon, $c$-quark and $b$-quark 
in the littlest Higgs model and the SUSY $E_6$ models in 
Fig.~\ref{fig:asymmetry}. 
The numbers of the asymmetries in each model are summarized 
in Table~\ref{asym}. 
It is remarkable that the asymmetries in the $\psi$ model are zero
because the extra U(1) charge assignments on the SM fermions are 
parity invariant. 
Beside on the $\psi$-model, the difference of predictions between 
the littlest Higgs model and the SUSY $E_6$ models are very clear in 
the $b$- and $c$-quark asymmetries. 
In the $b$-quark FB asymmetry, the littlest Higgs model predicts a
positive value while the SUSY $E_6$ model are negative one. 
Especially, it is noticeable that there is no $c$-quark asymmetry in 
the SUSY $E_6$ models though the littlest Higgs model gives a 40\% 
asymmetry.  
The reason why the FB asymmetry of $c$-quark vanishes in SUSY $E_6$ 
models is as follows. 
As shown in eq.~(\ref{eq_asym}), the FB asymmetry is given by the 
difference of the couplings between the left- and right-handed 
fermions to the $Z'$ boson. 
In SUSY $E_6$ models, both left- and right-handed $c$-quarks are
embedded in the same multiplet, $\mathbf{10}$ representation in SU(5), 
so that they have a common coupling which leads to no asymmetry.  
Fig.~\ref{fig:asymmetry} tells us that the measurements of $b$- and 
$c$-quark asymmetries in a few \% accuracy is enough to test 
if a $Z'$ boson is $A_H$ in the littlest Higgs model or one of 
the SUSY $E_6$ models. 

In summary, we have studied a possibility to test the littlest Higgs 
model and the SUSY $E_6$ models through $Z'$ production and decay 
at the $e^+e^-$ linear collider experiment. 
In our study, it is supposed that the LHC experiment discovers a $Z'$
boson before the linear collider experiment starts, and the $Z'$ boson 
is as light as $\sqrt{s}$ of the linear collider. 
Under this condition, we studied the $e^+ e^- \to f\bar{f}$ process 
at the $Z'$ pole. 
We find that, since the peak cross section of $e^+e^- \to \mu^+\mu^-$ in
the littlest Higgs model is roughly an order of magnitude larger than 
SUSY $E_6$ models in most of parameter space, we can rather easily 
test the models. 
However, if the mixing parameter $\theta'$ in the littlest Higgs model 
is close to a certain value, the peak
cross section decreases as small as one of the SUSY $E_6$ models. 
We show that the forward-backward asymmetry is independent
of the parameter $\theta'$  so that it is suitable to test the model
when measured cross section is close to one of the SUSY $E_6$ models. 
We find that discrepancies of the model predictions are very 
clear in $b$- and $c$-quark asymmetries. 
This is because that the couplings of quarks and leptons 
to $Z'$ boson in 
SUSY $E_6$ models are determined taking account of the SU(5) GUT 
symmetry while those in the littlest Higgs model are determined by 
different way. 
In particular, the $c$-quark FB asymmetry is predicted to be 40\% in 
the littlest Higgs model while zero in the SUSY $E_6$ models because, 
the left- and right-handed $c$-quarks have same couplings to the $Z'$
boson due to the SU(5) GUT symmetry. 
The measurements of FB-asymmetries for heavy quarks are, therefore, very 
useful to test if the $Z'$ boson is $A_H$ in the littlest Higgs model or 
one of the SUSY $E_6$ models. 

If the measurements of FB-asymmetries are consistent with the prediction
of the littlest Higgs model, one should determine the couplings of
fermion pairs to the $A_H$ boson for completeness. 
We briefly comment on this possibility before close our paper. 
From the data of $A_{\rm FB}^\mu,A_{\rm FB}^c,A_{\rm FB}^b$, we can
obtain the parameter $r_e,r_u,r_d$, respectively, where we assume that 
the couplings are universal for each generation. Taking account of the
$r_f$-parameters, the absolute values of the left- and right-handed 
couplings can be extracted from the partial decay width of 
$A_H \to f\bar{f}$ (\ref{width_a}). 
In order to fix the sign of the couplings, the interference effect
between $A_H$ and the SM $Z$ boson in the cross section of 
$e^+ e^- \to f \bar{f}$ at the off-resonance of $A_H$ should be measured 
very precisely.

\begin{table}
\begin{center}
\begin{tabular}{|c||c|c|c|c|c|} \hline
 & $A_H$ & $\chi$ & $\psi$ & $\eta$ & $\nu$ \\ \hline
$A_{\rm FB}^\mu$ & 0.27 & \hphantom{-} 0.48 & 0 
& \hphantom{-} 0.27 & \hphantom{-} 0.27 \\
$A_{\rm FB}^b$ & 0.27 & $-0.48$ & 0 & $-0.27$ & $-0.27$ \\
$A_{\rm FB}^c$ & 0.40 & 0 & 0 & 0 & 0 \\ \hline
\end{tabular}
\end{center}
\caption{Forward-backward asymmetry $A_{\rm FB}$ for 
muon, $b$ and $c$ quarks in LHM and SUSY $E_6$ models. 
}
\label{asym}
\end{table}

\begin{acknowledgments}
The work of G.C.C is supported in part by the Grant-in-Aid for Science 
Research, Ministry of Education, Science and Culture, Japan 
(No.13740149). 
\end{acknowledgments}

\end{document}